\documentclass[prl,superscriptaddress,preprint,showpacs]{revtex4}

\usepackage{amssymb}
\usepackage{graphicx}
\usepackage{dcolumn}
\usepackage{amsmath}

\begin{document}

\title{Field dependent effective masses in YbAl$_{3}$}
\author{T. Ebihara}
\affiliation{Shizuoka University, Shizuoka 422-8529, Japan}
\author{A. L. Cornelius}
\affiliation{University of Nevada, Las Vegas, NV 89154}
\author{J. M. Lawrence}
\affiliation{University of California, Irvine CA 92697}
\email{jmlawren@uci.edu}
\author{S. Uji}
\affiliation{National Institute for Materials Science, Tsukuba
305-0003, Japan}
\author{N. Harrison}
\affiliation{National High Magnetic Field Laboratory, Los Alamos
NM 87545}

\date{\today}

\begin{abstract}
We show for the intermediate valence compound YbAl$_{3}$ that the
high field (40 $\lesssim B \lesssim$ 60T) effective masses
measured by the de Haas-van Alphen experiment for field along the
$<111>$ direction are smaller by approximately a factor of two
than the low field masses. The field $B^{*} \sim$ 40T for this
reduction is much smaller than the Kondo field $B_{K} \sim
k_{B}T_{K}/\mu_{B}$ ($T_{K}\sim$ 670K) but is comparable to the
field $k_{B}T_{coh}/\mu_{B}$ where $T_{coh}\sim$ 40K is the
temperature for the onset of Fermi liquid coherence.  This
suggests that the field scale $B^{*}$ does not arise from 4$f$
polarization but is connected with the removal of the anomalies
that are known to occur in the Fermi liquid state of this
compound.
\end{abstract}


\maketitle

The de Haas-van Alphen (dHvA) frequencies measured for nonmagnetic
intermediate valence (IV) compounds such as CeSn$_{3}$ are
generally in good agreement with the predictions of the LDA band
theory that treats the 4$f$ electron as itinerant\cite{Hasegawa}.
While the LDA thus appears to give the geometry of the Fermi
surface correctly, it seriously underestimates the large measured
effective masses that arise in these strongly correlated systems.
For CeSn$_{3}$ some of this disagreement can be remedied by use of
the LDA+U approach\cite{Tanaka}.

For the IV compound YbAl$_{3}$ the LDA\cite{Ebihara} fails to
reproduce the measured Fermi surface.  This failure is connected
to the LDA prediction that the Yb 4$f$ level is fully occupied
(4$f^{14}$ or divalent Yb).   Experiments\cite{Cornelius} indicate
nonintegral valence of the Yb (4$f^{14-n_{f}}$ with $n_{f}=$
0.75). Excellent agreement with the experimental dHvA
frequencies\cite{Ebihara} of YbAl$_{3}$ has been obtained in an
LDA treatment where the energy of the 4$f$ level is constrained in
such a way as to permit nonintegral 4$f$ occupation. The divalence
that is incorrectly predicted by the LDA for other Yb compounds
has also been remedied\cite{Antonov} by use of LDA+U; clearly the
correlations affect the 4f level position as well as the effective
masses.

For heavy fermion compounds with small Kondo temperatures, the
application of a magnetic field larger than the Kondo field $B_{K}
\sim k_{B}T_{K}/\mu_{B}$ is known to cause a large reduction in
the measured effective masses\cite{Harrison}. This finds a natural
explanation in the theory of a Kondo impurity\cite{Schlottmann}
or of the Anderson lattice\cite{Wasserman}: when the Zeeman
splitting becomes larger than the Kondo temperature, the 4$f$
level polarizes; the resulting reduction of the spin degeneracy
inhibits the Kondo interaction that is responsible for the large
effective masses. In this paper we show that the effective masses
of key branches of the Fermi surface of YbAl$_{3}$ reduce by
approximately a factor of two in magnetic fields of order $B^{*}=$
40T. A field of this magnitude is substantially smaller than the
Kondo field $B_{K} \sim k_{B}T_{K}/\mu_{B}$  for this compound
($T_{K}=$ 670K\cite{Cornelius}) and hence the transition cannot
arise from 4$f$ polarization. The field energy $\mu_{B}B^{*}$
\textit{does}, however, correspond to the low energy scale $k_{B}
T_{coh}$ ($T_{coh}\sim$ 40K) for the onset of Fermi liquid
coherence in YbAl$_{3}$\cite{Ebihara, Cornelius}. We argue that
the reduction of the effective mass is connected with the removal
of the anomalous behavior\cite{Cornelius} that occurs for $T <
T_{coh}$ in this compound.

Single crystals of YbAl$_{3}$ were grown by the self-flux method,
with excess aluminum\cite{Ebihara}.  The crystals were aligned
using x-ray diffraction and reduced to the appropriate dimensions
using a spark cutter followed by etching in acid solutions.  The
low-field (13-17T) and intermediate field (17-19.5T) dHvA
measurements were performed using the field-modulation technique
at the National Institute for Materials Science in Tsukuba. The
low field measurements allowed for measurement as a function of
field angle relative to the high symmetry directions, and were
performed using a dilution refrigerator with a lower temperature
limit of 0.05K. The intermediate field measurements were performed
at fixed angle using a He-3 refrigerator. The high field ($B\leq$
57T) dHvA signal was measured using the pulsed-field technique
with counter-wound highly compensated pickup coils in the 60T
short-pulse magnet at the National High Magnetic Field Laboratory
in Los Alamos. The field angle was fixed for each sample, and the
temperature was regulated between 0.5-1.5K using a He-3
refrigerator.

\begin{figure}[e]
\includegraphics[width=3.3in]{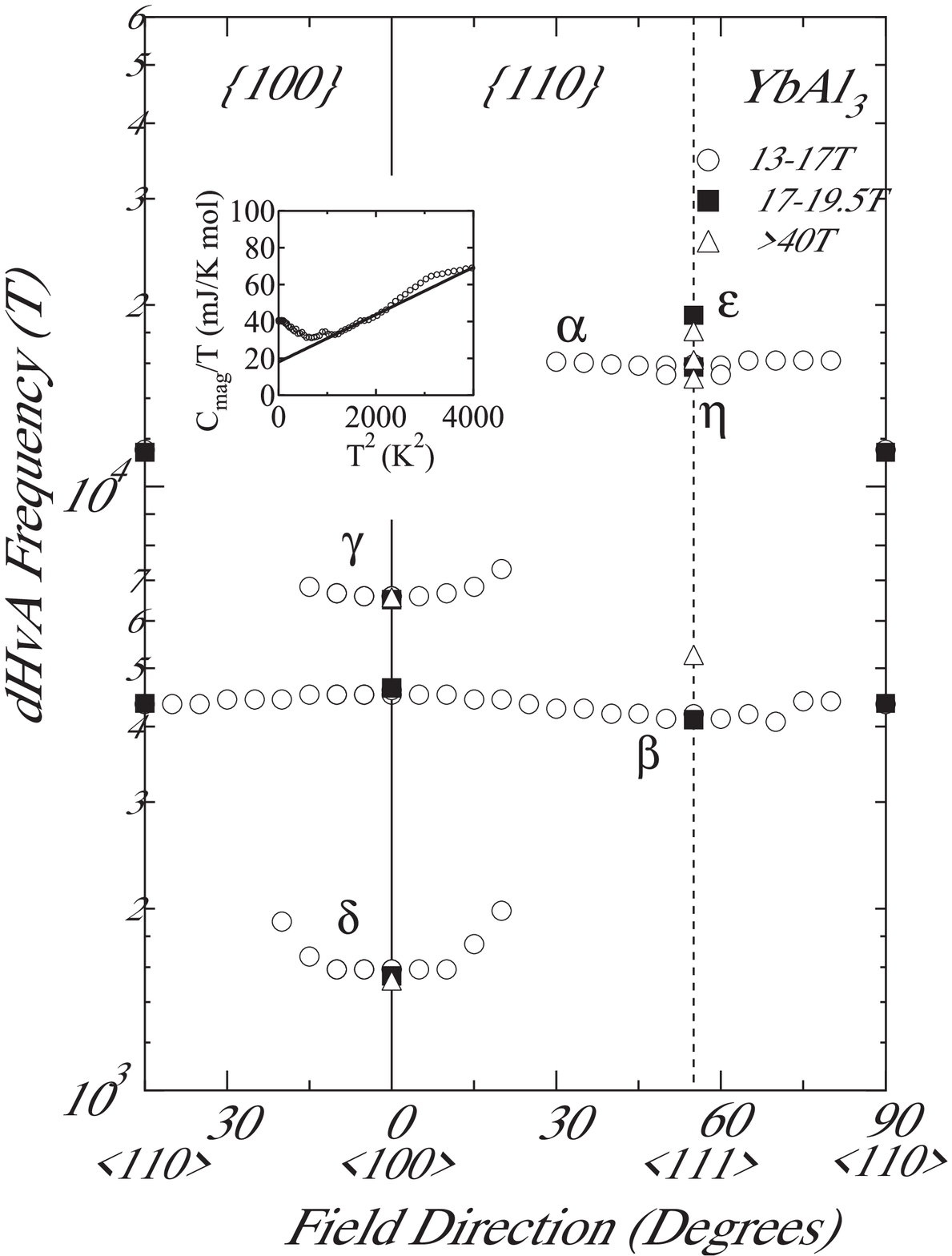}
\caption{ The de
Haas-van Alphen frequencies of YbAl$_{3}$ as a function of angle
with respect to the high symmetry directions. The different
symbols correspond to different field ranges for the measurement.
The inset shows the coefficient $C_{mag}/T$ of the magnetic
contribution to the specific heat as a function of $T^{2}$ to
demonstrate that the low field value is enhanced by a factor of
two with respect to the value extrapolated from higher
temperature.}
\end{figure}

\begin{figure}[e]
\includegraphics[width=3.3in]{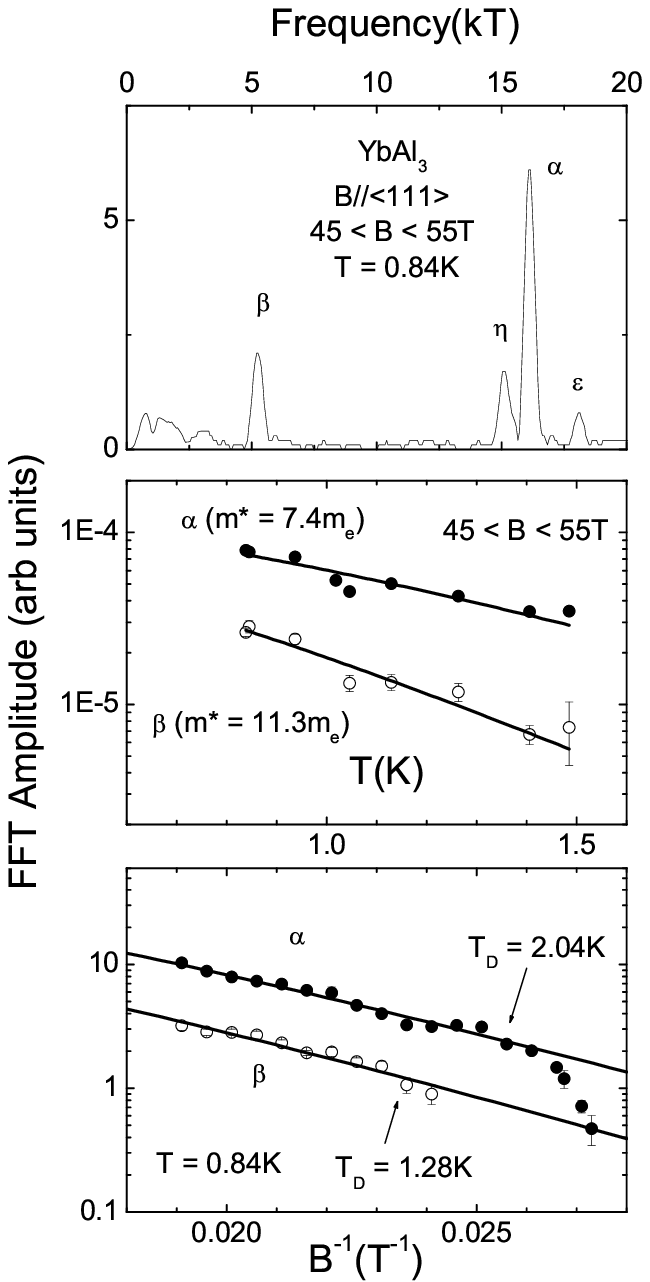}
\caption{ a) The fast Fourier transform of the de Haas-van Alphen
spectrum at 0.84K for field along the $<111>$ direction for fields
in the range 45 $\leq B \leq$ 55T.  b) A mass plot and c) a Dingle
plot for the $\alpha$ and $\beta$ branches for field along
$<111>$.  (See text for the fitting function.)}
\end{figure}

The dHvA frequencies for the low-field measurements are shown in
Fig. 1.  These reproduce the older work\cite{Ebihara} with one
nearly spherical high mass (23-29$m_{e}$) branch, denoted $\beta$,
and several branches that are confined to the vicinity of
particular field directions. The frequencies for intermediate
field are essentially unchanged from those at low field.  In Fig.
2a we show the fast Fourier transform (FFT) of the dHvA signal at
high field (45 $<B<$ 55T) at 0.84K and for field along the $<111>$
direction. Four peaks are resolved, which correspond in frequency
to the $\alpha$, $\beta$, $\epsilon$ and $\eta$ branches observed
at low field\cite{Ebihara}.  The high field frequencies are equal,
within a few percent, to those observed at low field (Fig. 1) with
the exception of the frequency of the $\beta$ branch, which is
25\% larger than at low field.

We analyzed the high field measurements for each field range and
temperature by fitting each FFT peak to the sum of a Gaussian and
a second-order polynomial background function. We fit the Gaussian
amplitudes to the function
 \begin{equation}
 A(B,T)\varpropto \left(
\frac{F^{2}T}{B^{5/2}}\right) \left( \frac{\partial B}{\partial
t}\right) \frac{\exp [14.7m^{\ast }T_{D}/B]}{\sinh [14.7m^{\ast}
T/B]}
\end{equation}
where $F$ is the dHvA frequency, $m^{*}$ is the effective mass and
$T_{D}$ is the Dingle temperature. This formula is appropriate for
pulsed field data\cite{Harrison}; a related expression appropriate
for d.c. fields was used for the low and intermediate field data.
By fitting at fixed field as a function of temperature we
determined the effective mass for each branch and field range.
Examples for the $\alpha$ and $\beta$ branches are shown in Fig.
2b. By fitting at fixed temperature as a function of field, we
determine the Dingle temperature (Fig. 2c).

\begin{figure}[e]
\includegraphics[width=3.3in]{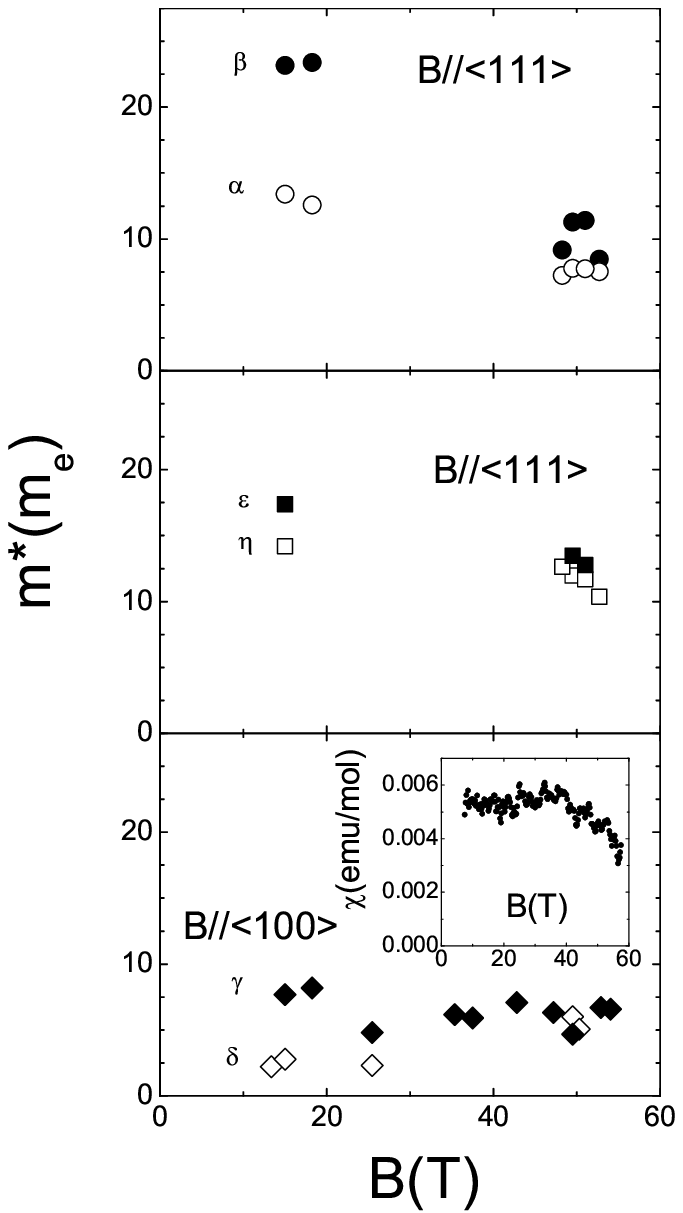}
\caption{ The dHvA masses in units of the free electron mass
$m_{e}$ for different branches of the Fermi surface of YbAl$_{3}$
as a function of magnetic field for a) and b) field along the
$<111>$ direction and c) field along the $<100>$ direction. A
large reduction of the $\alpha$ and $\beta$ branches is observed
at high field. Inset: The field dependence of the magnetic
susceptibility, showing that $\chi(B)$ decreases for $B
>$ 40T. }
\end{figure}

The masses determined from this analysis are plotted as a function
of field in Fig. 3.  The plot includes the low and intermediate
field values, and data from two different samples for each field
orientation at high field.  The scatter at high field gives a good
indication of both the reproducibility and the statistical error.
Our key result is that the masses of the $\alpha$ and $\beta$
branches for field along the $<111>$ direction decrease by
approximately a factor of two above 40T.  Somewhat smaller
reductions are observed for the $\eta$ and $\epsilon$ branches.
For field along the $<100>$ direction there is a more modest
decrease of the $\gamma$ mass and the mass of the $\delta$ branch
appears to \textit{increase} with field.

We did not observe a dHvA signal for the $\beta$ branch for field
along either the $<100>$ or the $<110>$ directions.  The low field
masses for this branch for these directions are 28.9 and
26.3$m_{e}$ respectively.  From Equation 1 it is clear that large
masses suppress the dHvA amplitude except at the lowest
temperatures.  In the pulsed field measurement, using a He-3
refrigerator, self-heating due to eddy currents limited our
measurements to $T >$ 0.5K.  In this situation, it is very
difficult to detect dHvA signals for branches with $m^{*}$ much
greater than 15$m_{e}$.  Our data is thus consistent with a
reduction of the mass of the $\beta$ branch for field in these
directions, with a lower limit on the high field mass of order
15$m_{e}$.

For similar reasons it was difficult to observe dHvA signals for
the high mass branches for fields below about 35T.  The field
$B^{*}$ for the transition to reduced masses is constrained by the
dHvA data (Fig. 3) to lie somewhere in the interval 20 $<B^{*}<$
45T. The susceptibility (Fig. 3, inset) is constant up to 40T, and
decreases at higher field, suggesting that $B^{*}\sim$ 40T.  The
field dependence of the amplitude of the $\alpha$ branch fits the
predicted Dingle curve for $B >$ 38T (Fig. 2c) which also suggests
a transition field of this magnitude.

We have recently shown\cite{Ebihara, Cornelius} that $T_{coh}\sim$
40K is the temperature below which the electrical resistivity
exhibits the $T^{2}$ behavior expected for a Fermi liquid. The low
temperature optical conductivity of YbAl$_{3}$
exhibits\cite{Okamura} both a very narrow Drude resonance and a
mid-infrared peak that is associated with optical transitions
across the renormalized hybridization gap. The Drude peak begins
to broaden and the mid-IR peak begins to be suppressed above 40K.
Hence, the optical conductivity indicates that 40K is the
temperature scale for full establishment of the renormalized band
structure.  In a 1$/N_{J}$ -expansion treatment\cite{Kuroda} of
the Anderson lattice for $N_{J}=2J+1=$ 6, the coherence
temperature (defined here also as the temperature where the
renormalized Fermi liquid is fully established) was found to
satisfy $T_{coh}\approx T_{K}/$10, which is approximately true for
YbAl$_{3}$.  In this treatment, a low temperature anomaly in the
susceptibility was predicted for $T < T_{coh}$ when the total
electron density $n_{c}+n_{f}$ was set to a value (1.9) that is
close to the value appropriate for a Kondo insulator, i.e. in the
limit of low conduction hole density. A recent slave Boson
treatment\cite{Georges} of the Anderson lattice in the limit of
low conduction \textit{electron} density also shows anomalies in
the susceptibility and specific heat for $T < T_{coh}$. Hence our
recent experimental results for YbAl$_{3}$ \cite{Cornelius} which
show both a suceptibility anomaly and a specific heat anomaly
(Fig. 1, inset) below 40K may be explained by the Anderson lattice
in the limit of low conduction electron or hole density.

The results reported here show that above a transition field
$B^{*}\sim$ 40T the dHvA masses for key branches of Fermi surface
decrease by approximately a factor of two.  The specific heat
anomaly (Fig. 1 inset) is consistent with this, in that the low
temperature value is twice as large as the value extrapolated
from higher temperature; and we predict that the specific heat
coefficient will decrease by a factor of two above 40T.  As
discussed above, we expect the transition field for polarization
of the 4$f$ level in the Anderson lattice to satisfy
$gJ\mu_{B}B_{K}\approx k_{B}T_{K}$\cite{Schlottmann, Wasserman,
Ono}. For YbAl$_{3}$ where $g=$ 8/7, $J=$ 7/2 and $T_{K}=$ 670K,
this formula gives $B_{K}=$ 250K, much larger than the observed
value of $B^{*}$. The theory also predicts that the
susceptibility will increase as the field increases in the
vicinity of $B_{K}$; the 4$f$ polarization can be thought of as a
continuous metamagnetic transition\cite{Ono}. However, the
susceptibility of YbAl$_{3}$ \textit{decreases} at $B^{*}$.
Finally, for systems such as CeRu$_{2}$Si$_{2}$ where the
metamagnetic transition is observed experimentally, the Fermi
surface alters at the transition because the polarization
corresponds to a localization of the 4$f$ electron which causes
it to drop out of the Fermi volume\cite{Onuki}.  For YbAl$_{3}$,
however, the transition has only a small effect on the Fermi
surface geometry, with most frequencies unchanged but with a 25\%
increase in the frequency of the $\beta$ branch along $<111>$
which is suggestive of an \textit{increase} in the Fermi volume.
These facts suggest that the reduction of the masses is not
connected with a simple polarization of the 4$f$ level.

The transition field energy \textit{is}, however, of order of
$k_{B}T_{coh}$, which suggests that the field dependence of the
masses is intimately related to the existence of the low
temperature anomalies. For $B > B^{*}$ the susceptibility anomaly
is suppressed and the temperature dependence becomes qualitatively
similar to that of an Anderson impurity\cite{Cornelius}.  The
effective masses reduce to values that are still moderately large,
comparable indeed to those of CeSn$_{3}$\cite{Tanaka}, without
much change in the Fermi surface geometry.  It is as though the
transition turns the compound into a non-anomalous intermediate
valence compound, with moderately renormalized bands, but without
4$f$ polarization. It remains to be seen whether the theory of the
Anderson lattice can explain these effects and clarify the
difference between anomalous and non-anomalous IV compounds.

Work by T. E. in the U.S. was supported by the Research Abroad
Program of the Japanese Ministry of Education; work in Japan was
supported in part by the Inoue Science Promotion Foundation and in
part by Corning Japan. Work at UNLV was supported by DOE
EPSCoR-State/National Laboratory Partnership Award
DE-FG02-00ER45835 and DOE Cooperative Agreement DE-FC08-98NV1341.
Work by J.L. in Japan was supported by the Japanese Society for
the Promotion of Science; work at UC Irvine was supported by UCDRD
funds provided by the University of California for the conduct of
discretionary research by the Los Alamos National Laboratory and
by the UC/LANL Personnel Assignment Program.  Work at the National
High Magnetic Field Laboratory, Los Alamos Facility, was performed
under the auspices of the National Science Foundation, the State
of Florida and the Department of Energy.


\begin{references}

\bibitem{Hasegawa}  A. Hasegawa and H. Yamagami, Progress of Theoretical Physics
Supplement {\bf 108}, 27 (1992).

\bibitem{Tanaka} S. Tanaka, H. Harima and A. Yanase, Journ. Mag.
Mag. Mat. {\bf 177-181}, 329 (1998).

\bibitem{Ebihara} T. Ebihara, Y. Inada, M.Murakawa, S. Uji, C. Terakura, T. Terashima,
E. Yamamoto, Y. Haga, Y. Onuki and H. Harima, Journ. Phys. Soc.
Japan {\bf 69}, 895 (2000).

\bibitem{Cornelius} A. L. Cornelius, J. M. Lawrence, T. Ebihara,
P.S. Riseborough, C. H. Booth, M. F. Hundley, P. G. Pagliuso, J.
L. Sarrao, J. D. Thompson, M. H. Jung, A. H. Lacerda and G. H.
Kwei, Phys. Rev. Lett. {\bf 88}, 117201 (2002).


\bibitem{Antonov} V. N. Antonov, M. Galli, F. Marabelli, A. N.
Yaresko, A. Ya. Perlov and E. Bauer, Phys. Rev. B {\bf 62}, 1742
(2000).

\bibitem{Harrison}N. Harrison, P. Meeson, P.-A. Probst and M.
Springford, J. Phys.: Condens. Matter {\bf 5}, 7435 (1993).

\bibitem{Schlottmann} P. Schlottmann, Phys. Rev. B {\bf 35}, 5279
(1987).

\bibitem{Wasserman} A. Wasserman, M. Springford and A. C. Hewson,
J. Phys.: Condens. Matter {\bf 1}, 2669 (1989).

\bibitem{Okamura} H. Okamura, T. Ebihara and T. Nanba, Acta
Physica Polonica, to be published (cond-mat/0208006); and
unpublished data.

\bibitem{Kuroda} Y. $\bar{\rm O}$no, T. Matsuura and Y. Kuroda, Journ. Phys.
Soc. Japan {\bf 60}, 3475 (1991).

\bibitem{Georges} S. Burdin, A. Georges and D. R. Grempel, Phys.
Rev. Lett. {\bf 85}, 1048 (2000).

\bibitem{Ono}  Y. $\bar{\rm O}$no, Journ. Phys. Soc. Japan {\bf 65}, 19
(1996).

\bibitem{Onuki}  Y. $\bar{\rm O}$nuki, R. Settai and H. Aoki,
Physica B {\bf 223\&224}, 141 (1996).





\end{references}
\end{document}